\documentclass{elsart}
\usepackage{graphicx}
\usepackage{array,tabularx}
\newcommand{\pt}{$p_T$}
\newcommand{\Ks}{$K^{0}_{S}$}
\newcommand{\La}{$\Lambda$}
\newcommand{\aLa}{$\overline{\Lambda}$}
\newcommand{\pp}{$p+p$}
\usepackage{epsfig}

 \usepackage{amssymb}

\usepackage{lineno}
\linenumbers
\def\Journal#1#2#3#4{{#1} {\bf #2 }(#3) #4}

\def\PRL{ Phys. Rev. Lett.}

\def\PRC{{ Phys. Rev.} C}

\begin{document}

\begin{frontmatter}
 %thanks[label1]{}
 %\author{Name\corauthref{cor1}\thanksref{label2}}
 %\ead{email address}
 %\thanks[label2]{}
 %\corauth[cor1]{}
%\author[label1,label2]{}
%\address[label1]{}
%\address[label2]{}
%\address{Address\thanksref{label3}}
 %\thanks[label3]{}

%\title{Improving the $dE/dx$ calibration of the STAR TPC for the high-$p_T$ hadron identification}
%\title{Extending particle identification to the high-$p_T$ using the STAR Electro-Magnetic-Calorimeter as hadron trigger}
\title{Hadronic Trigger using electromagnetic calorimeter and particle identification at high-$p_T$ with the STAR Detector}

\author[ustc]{Hongyu Da},
\author[ustc,bnl]{Xiangli Cui},
\author [bnl]{Gene Van Buren},
%\author[buk]{Lee Barnby},
\author[bnl]{J.C. Dunlop}, %James Dunlop
\author [lbl]{Xin Dong},
%\author[bnl]{Patricia Fachini},
%\author[bnl]{Ron Longacre},
\author [bnl]{Lijuan Ruan},
\author[houston]{Anthony Timmins},
\author [ustc]{Zebo Tang},
\author [ustc]{Xiaolian Wang}
\author [ustc]{Yichun Xu}\ead{xuyichun@mail.ustc.edu.cn},
\author [bnl]{Zhangbu Xu}
\address [ustc]{Department of Modern Physics, University of Science and Technology of China, Hefei, Anhui, China, 230026}
\address [bnl]{Department of Physics, Brookhaven National Laboratory, Upton, NY 11973, USA}
%\address[uw]{Nuclear Physics Laboratory, University of Washington, Box 354290, Seattle, WA 98195-4290, USA}
\address [lbl]{Lawrence Berkeley National Laboratory, 1 Cyclotron Road, Berkeley, CA 94720, USA}
\address [houston]{Physics Department, 617 Science \& Research Building 1, Houston, TX 77204 }
%\address[buk]{University of Birmingham, Birmingham, United Kingdom}

\begin{abstract}
% Text of abstract

We derive a new method to improve the statistics of identified
particles at high transverse momentum (\pt) using online-triggered
events by the STAR Barrel Electro-Magnetic-Calorimeter (BEMC) detector.
The BEMC is used to select charged hadrons ($\pi^{\pm},K^{\pm}$ and
$p$($\bar{p}$)) via hadronic shower energy depostied in the BEMC.
With this trigger, the statistics of
the high \pt~particles are significantly enhanced (by a factor
of up to $\sim$100 for STAR) with trigger efficiency up to 20\%. In addition, resonant
states ($\rho^0$, $K^{\star}$) and weak-decay V0s (\Ks~and \La(\aLa))
can be constructed by selecting the BEMC-trigger hadron as one of
the decay daughters. We also show that the trigger efficiency can be
obtained reliably in simulation and data-driven approaches.

\end{abstract}

\end{frontmatter}

% main text
\section{Introduction}
%physical goals(??)=>detector(TPC+EMC)

One of the main physics goals of the Relativistic Heavy Ion Collider
(RHIC) with the experiment of the Solenoidal Tracker at RHIC (STAR)~\cite{STAR}
in recent (and future) years is to study the properties of the
Quark-Gluon-Plasma (QGP) created in the heavy ion
collisions~\cite{starwhitepaper}. An important probe is to use the
identified high-\pt~hadrons to study the color charge effect of
parton energy loss in heavy ion
collisions\cite{starPIDpapers,starppPID,SQM09,HQ08,QM09,starppStrange}.
At RHIC, the luminosity is usually much higher than the detector and
data acquisition capability. RHIC delivers \pp~collision rates of
several $MHz$ while the STAR Time Projection Chamber (TPC) readout
is around one $kHz$. Trigger detectors are used to implement an
online selection of events of interest to the program. An example of
such a detector is the electromagnetic calorimeter (EMC) used to
select events with high energy deposit from an electromagnetic
shower in the detector. This can enhance the event sample with high
energy neutral pions or energetic jets with significant
electromagnetic components ($\pi^{0}$ or $\gamma$). However, STAR
has no hadronic calorimeter to select the final state charged
hadrons although those hadrons do leave ionization tracks behind in
the TPC or other tracking detectors. To date, charged hadron spectra
in \pp~and $A+A$ collisions at RHIC have only been obtained using a
minimum-bias trigger and their upper reach in~\pt~is severely
limited by rapidly falling statistics at high-\pt~from such an
all-inclusive trigger. For example, the charged pion spectra are so
far only measured at RHIC to $p_T\simeq10$ $GeV/c$ in
\pp~collisions, while the $\pi^{0}$ spectra reach $p_T\simeq20$
$GeV/c$. On the other hand, the STAR EMC contains about one hadronic
interaction length of material and can perform online trigger
selection of events based on energy deposition in finely segmented
towers. Charged hadrons do interact and produce showers with a
significant amount of energy in the EMC at a lower efficiency.

In this paper, we present a study of hadronic trigger efficiency
from the STAR Barrel EMC (BEMC)~\cite{BEMC}. Different BEMC
energy thresholds and BEMC patch sizes are used in this study. The
inclusive charged hadron spectra are selected from the away-side
opposite the struck calorimeter tower (or jet patch). PYTHIA
simulations are performed to correct for the trigger effect. In
addition to the enhancement of single hadron yields at high
momentum, these triggers also allow us to construct resonances and
weak-decay particles from their charged hadron daughters by
requiring that one of the daughters produce a BEMC signal above the
trigger threshold. The trigger effect on resonance and V0
reconstructions is corrected using the experimental data. This
approach avoids a demanding simulation of the details of the
detector and trigger performances on the struck EMC tower. These
not only extend the measurements of identified hadron
spectra to much higher momentum, but also provide crucial
consistency checks among different measurements over the same momentum
range: $\pi^{0}$ vs $\pi^{\pm}$, $K^{0}_{_{S}}$ vs $K^{\pm}$. The
current manuscript provides the technical details of the triggers,
analyses, correction and systematics while the scientific results
have been discussed in Ref.~\cite{jetchemistry}.

\section{Experimental Setup and Data Analysis}
\subsection{Detectors and Datasets}

The data used for this study were collected with the STAR
Experiment in the year 2005
requiring the minimum-bias trigger condition plus energy deposition
in the BEMC
detector for $p+p$ collisions.
For this data, the total BEMC coverage is 0 $< \eta <$ 1 and
0 $< \phi \leq 2\pi$ $rad$. Each calorimeter tower covers
$\Delta\eta\times\Delta\phi$ = 0.05 $\times$ 0.05 $rad$ in pseudo-rapidity
($\eta$), and azimuthal angle ($\phi$). The online energy deposition
triggers utilize either a single BEMC tower (high-tower trigger, HT)
or a contiguous $\Delta\eta\times\Delta\phi$ =
$1\times1$ $rad$ region (jet patch trigger, JP) of the
BEMC~\cite{ppJet}. A total of
5.6 million JP events with transverse energy $E_{T} > $ 6.4 $GeV$ are
used for $\pi^{\pm}$, $K^{\pm}$, and $p(\bar{p})$ analyses. To
reduce trigger biases and to avoid the demandingly precise simulation
of hadronic showers and the detector trigger response, only away-side
particles (at azimuthal angles $90^{\circ}\!-\!270^{\circ}$ from the
JP trigger) are used in the analyses of the inclusive single
charged hadron spectra. The high-tower trigger condition
requires the energy of a single calorimeter tower to be at least 2.6
$GeV$ (HT1) or 3.5 $GeV$ (HT2)~\cite{Trigger,pi0eta}. In total, 5.1
million HT1 and 3.4 million HT2 events were collected from
0.65 $pb^{-1}$ and 2.83 $pb^{-1}$ integrated sampled
luminosity of proton beams. These datasets are used for
$K_S^{0}\rightarrow\pi^{+}+\pi^{-}$,
$\bar{\Lambda}\rightarrow\bar{p}+\pi^+$ and
$\rho^{0}\rightarrow\pi^{+}+\pi^{-}$ reconstruction by requiring
that one of the daughter pions or antiproton triggered the high
tower.

The TPC covers 0 $< \phi \leq $2$\pi$ $rad$ and $|\eta|\leq$ 1.3 with up
to 45 reconstructed hit points to serve as STAR's main tracking
detector. It measures ionization energy loss ($dE/dx$) and momentum
(via curvature) of tracks in a 0.5 $T$ solenoidal magnetic field,
which together can provide particle identification, including along
the relativistic rise at high momentum~\cite{TPC}. Topology of the
daughter tracks from high-\pt~\Ks, \La(\aLa), $\rho^{0}$,
$K^{\star}$, $D^{0}$ and other resonances can be reconstructed
through their hadronic decay into at least one high-\pt~charged
hadron. In all of the analyses discussed here, the collision vertex
is required to be within 100 $cm$ of the TPC center. More details of
hadron identification at high-\pt~\cite{dEdxCali,Ming,Bichsel}
 and topological V0 reconstruction of weak-decay particles~\cite{starppStrange} can be
found in the references.

\subsection{Single charged hadron spectra from triggered away-side jets}

In addition to the required azimuthal angle between the track and
the JP trigger center to be $|\Delta\phi| \ge \pi/2$ $rad$ in the analyses
of charged hadron spectra of $\pi^{\pm}$, $K^{\pm}$ and
$p$($\bar{p}$), the TPC tracks are selected based on: $|\eta| < 0.5$,
distance of closest approach of the track helix projection to the
collision vertex to be within 1.0 $cm$,  number of TPC hits
to be at least 25 and at least
52\% of the maximum possible hits, and the TPC hits involved in the
$dE/dx$ calculation after truncation to be at least 15.  In each
\pt~bin, the normalized $dE/dx$,
n$\sigma_{\pi}$~\cite{dEdxCali,Ming,Bichsel,lndEdx} distributions of
positively and negatively charged particles are histogrammed. The
detailed method of calibration and extraction of raw counts of the
individual identified hadrons from the same data sample has been
previously published~\cite{dEdxCali}.

\begin{figure}
\centering
\includegraphics[width=0.49\textwidth]{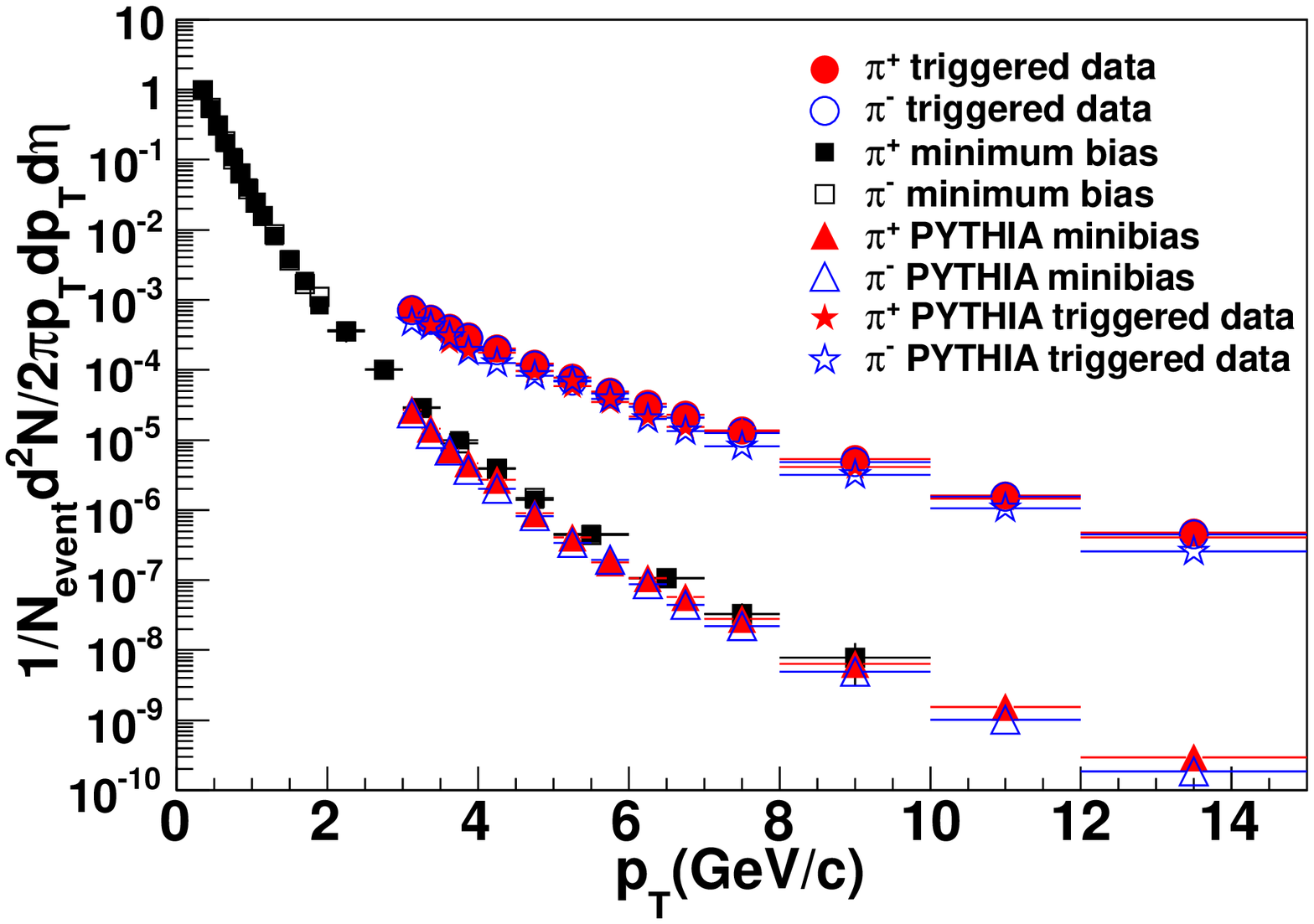} %pionSpectra_unCorr.eps}
\includegraphics[width=0.49\textwidth]{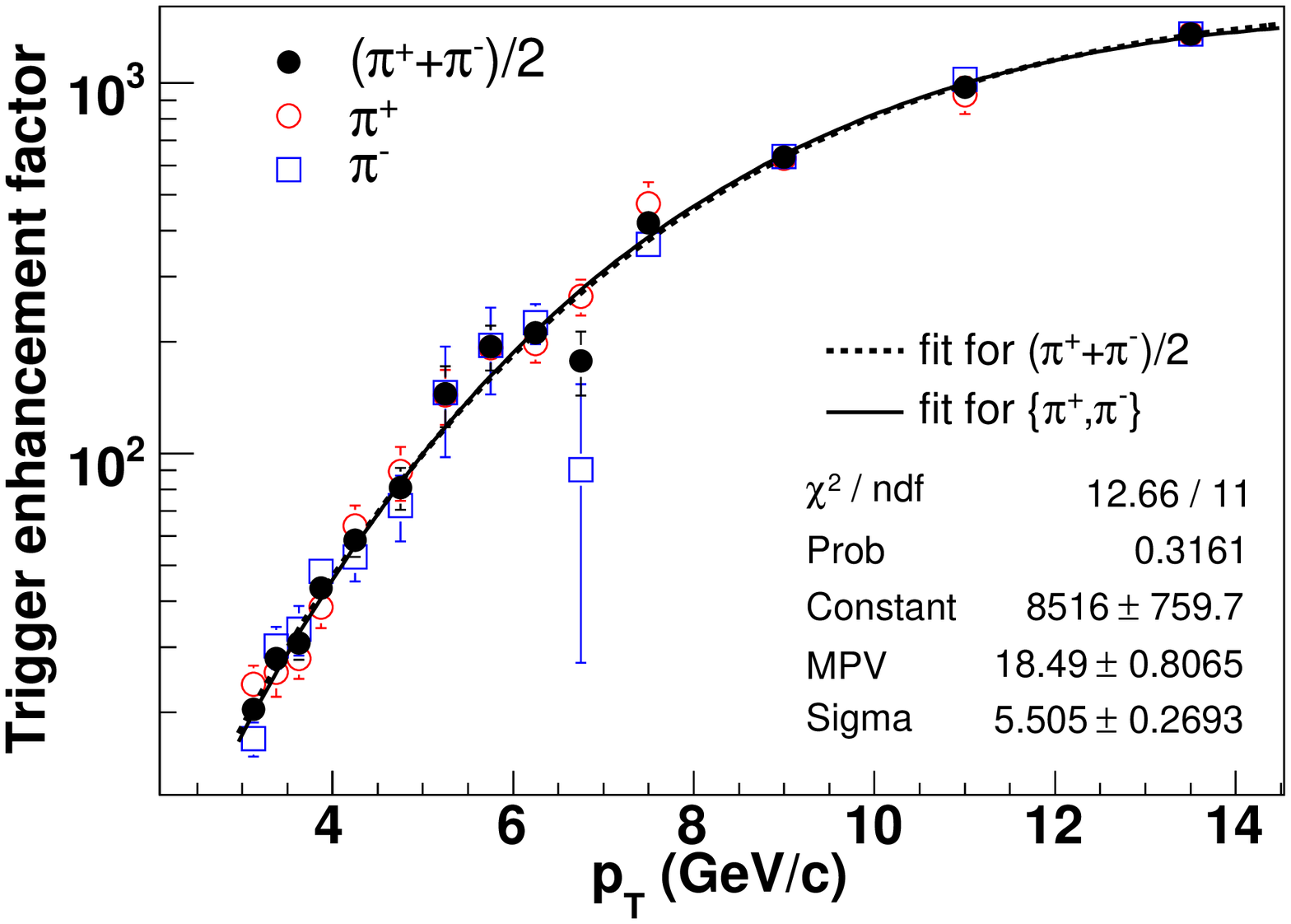}
\caption{The left panel shows pion spectra in minimum-bias and
BEMC-trigger events from both measurements and the PYTHIA+GEANT
simulation. Triggered enhancement from the simulations versus
\pt~distribution is shown on the right panel.}
\label{trgEnhancement}
\end{figure}

The JP triggers enhance the statistics greatly, but also require
additional corrections with normalization and momentum-dependent
efficiency. Figure~\ref{trgEnhancement} shows raw charged pion
spectra in the BEMC-trigger events compared to the published results
(squares) in minimum-bias events~\cite{starppPID}. This demonstrates
that charged pions in the BEMC-trigger data sample are enriched by
an order of magnitude at low $p_T$ ($\simeq 3$ $GeV/c$) and by three
orders of magnitude at high $p_T$ (${}^{>}_{\sim}10$ $GeV/c$). To
correct for this trigger effect, PYTHIA events are embedded into the
STAR detector geometry in GEANT, which can simulate the realistic
response of the STAR detector. The Monte Carlo simulation is based
on PYTHIA version 6.205~\cite{Sjostrand} with CDF Tune A settings
~\cite{CDFTune}. The same simulation setup has been used in other
jet related analyses~\cite{ppJet}. In order to fully cover the
falling power-law spectrum in \pt~ of reconstructed particles with
sufficient statistics, the data samples are generated according to
the initial parton $p_T$ (in units of $GeV/c$) intervals (0,2),
(2,3), (3,4), (4,5), (5,7), (7,9), (9,11), (11,15), (15,25), (25,35)
and ($>$35). The spectra are weighted by the cross-sections in each
parton $p_T$ range. Table~\ref{PYTHIA_parton} shows the absolute
cross-sections ($\sigma_i$) for $p+p$ collisions which generate the
partons at each given $p_T$ interval and the number of events
($N_i$) going through the full simulation chain. The obtained hadron
spectra have to be weighted with the factor (proportional to
$\sigma_i/N_i$) according to their originating partons.

\begin{figure}
\centering
\includegraphics[width=0.49\textwidth]{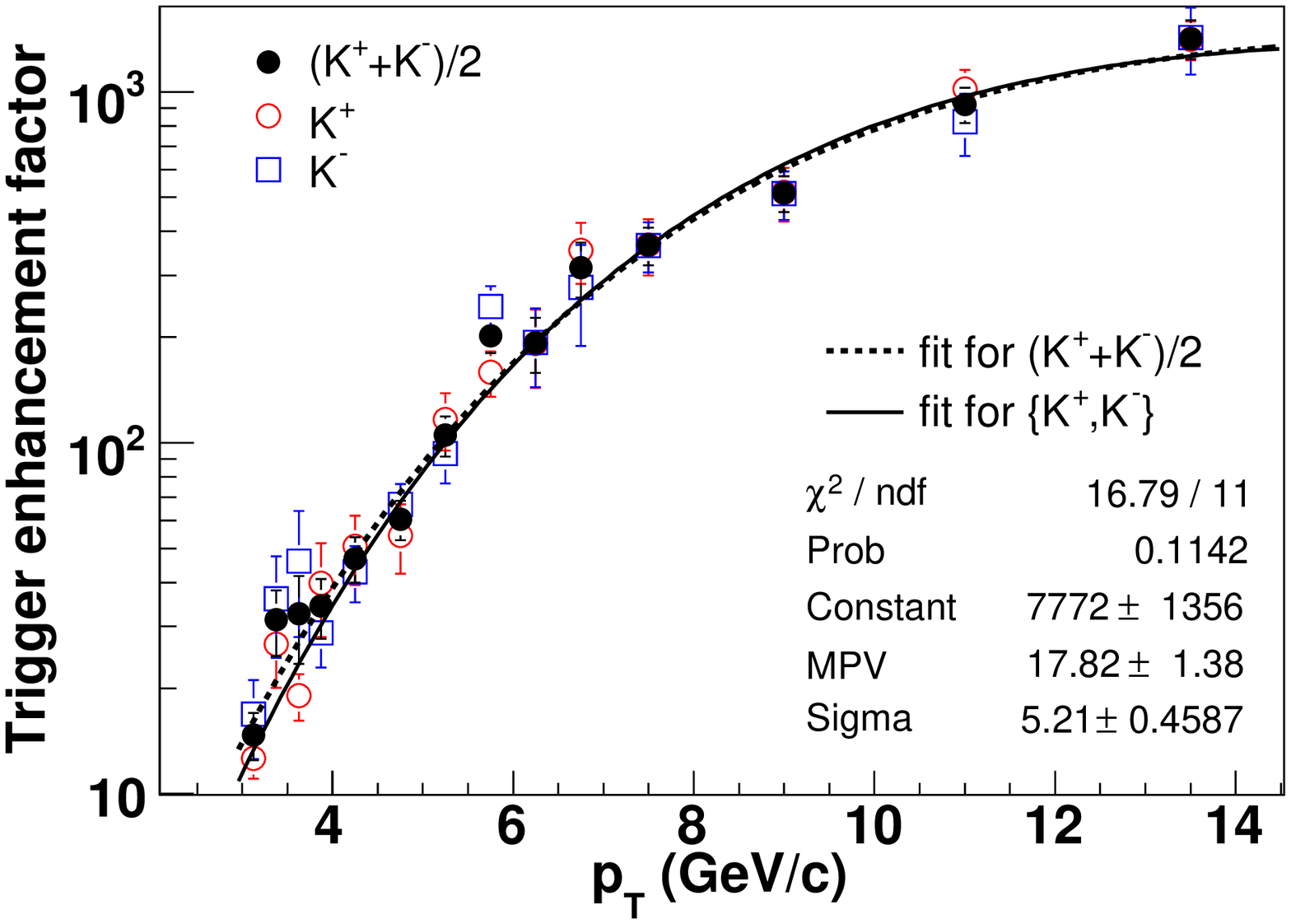}
\includegraphics[width=0.49\textwidth]{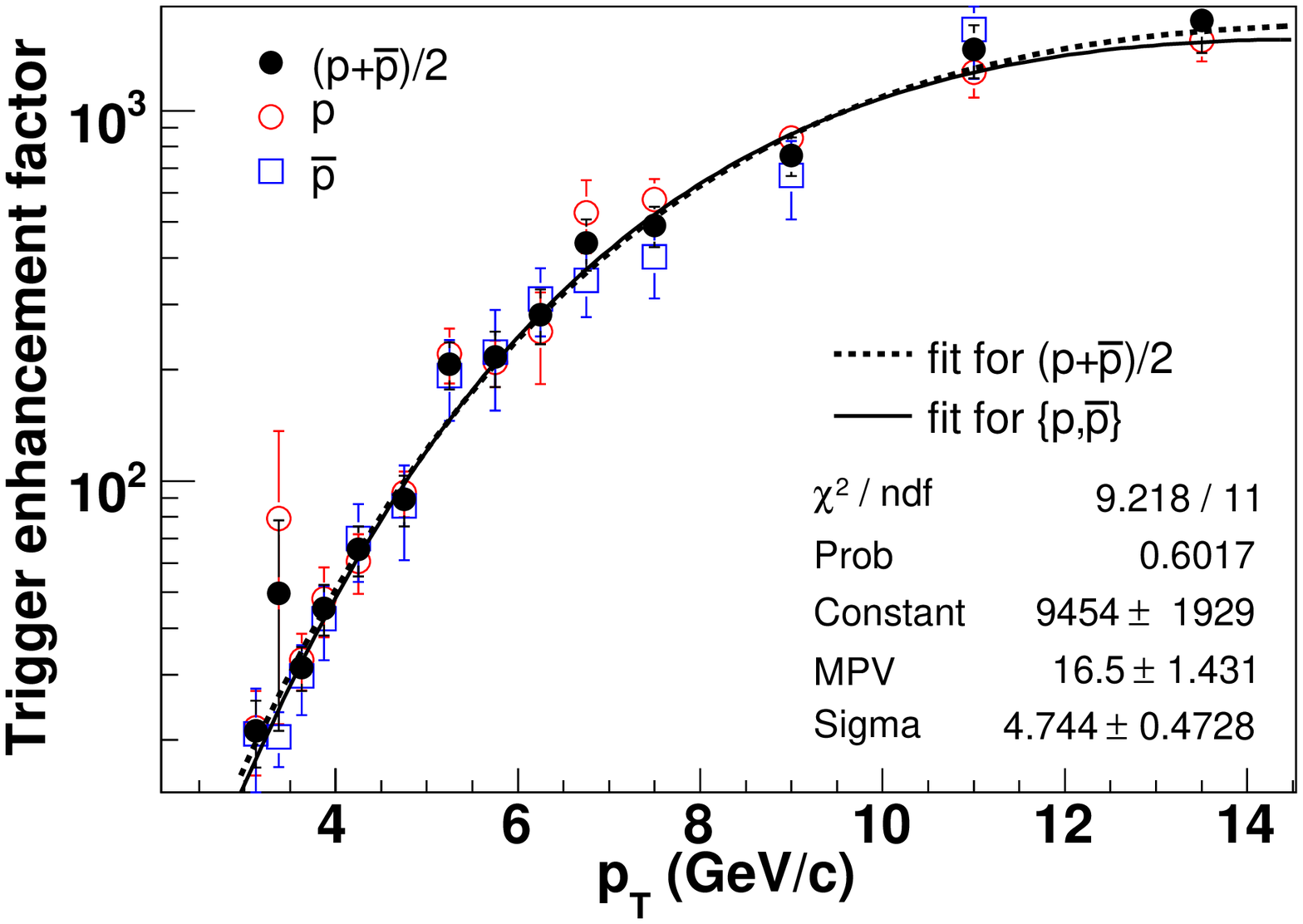}
\caption{Trigger enhancement factor distribution for kaon and proton
are shown on the left and right panels respectively.}
\label{trgEnhancementKP}
\end{figure}

\begin{table}
\begin{center}
\caption{Parton \pt~interval, the corresponding absolute cross-section and the number of
events generated in PYTHIA through the STAR simulation and reconstruction chain.}\label{PYTHIA_parton}
\begin{tabular}{|c|c|c|c|c|p{6.0cm}}
\hline \hline
parton \pt~($GeV/c$) &Cross-section ($mb$) &Number of events\\
\hline (0,2)  &18.2   &339083 \\
\hline
(2,3)  &8.11    &507996\\
\hline
(3,4)  &1.30   &400629\\
\hline
(4,5)  &0.314  &600980\\
\hline
(5,7)  &1.36e-1   &431000 \\
\hline
(7,9)   &2.31e-2    &412000\\
\hline
(9,11)  &5.51e-3    &416000\\
\hline
(11,15) &2.22e-3   &416000\\
\hline
(15,25) &3.89e-4  &408000\\
\hline
(25,35) &1.02e-5   &380000\\
\hline
( $>$ 35) &5.30e-7     &100000\\
\hline \hline
\end{tabular}
\end{center}
\end{table}

The simulation includes detector response to the signal, electronic
readout, and detector and background noise when particles propagate
through the detector. The BEMC-trigger configuration and thresholds
are then applied in the same way as in the real events from
experiment. The resulting charged pion spectra from these
simulations are shown on the left panel in
Figure~\ref{trgEnhancement} for the minimum-bias-trigger and the
BEMC-trigger events. The enhancement of charged pions can be
calculated by dividing the BEMC-trigger spectra by the
minimum-bias-trigger spectra from these PYTHIA simulations. The
right panel in Figure~\ref{trgEnhancement} shows the enhancement
factor as a function of transverse momentum. Similarly, the trigger
enhancement factors for kaon and proton are calculated and presented
in the left and right panels in Figure~\ref{trgEnhancementKP}. These
factors are then applied to the raw spectra to obtain the inclusive
invariant differential cross-section of the charged hadrons in $p+p$
collisions as presented in Ref.~\cite{jetchemistry}.

Since the correction for the JP trigger relies entirely on the PYTHIA
event generator through the STAR detector simulation chain, concerns
were raised whether the PYTHIA event generator simulates the jets
and underlying event structure correctly and whether the detector
simulation reproduces the JP trigger truthfully. To quanitfy this, several triggers
with different jet-patch and high-tower energy thresholds have been
used to study the systematic differences among the spectra after all
the corrections have been applied, providing an estimate of the systematic
uncertainty due to these effects. These are the largest contributions to the
overall systematic uncertainties, especially at
intermediate \pt~\cite{QM09,jetchemistry}.

Studies have shown that the underlying event
structure~\cite{Caines:2011zz} and the jet spectra~\cite{ppJet}
match well between data and PYTHIA. The results also show that the
averaged $\pi^{\pm}$ spectrum obtained from this method are
consistent with the $\pi^{0}$ spectra from both STAR~\cite{pi0eta}
and PHENIX~\cite{Adare:2008qa} to within $10\%$. The $\pi^{0}$
spectra were obtained with a completely different trigger scheme:
one of the photons from $\pi^{0}$ decay has to be reconstructed from
the BEMC high-tower, which triggers the event. Therefore, the
$\pi^{0}$ trigger is not affected by the event structure but depends
on the simulation of the photon response and trigger efficiency. In
the following sections, we describe a similar method to reconstruct
resonances and V0 decays by using a BEMC high-tower as a hadronic
trigger on one of the charged hadronic daughters. The trigger
efficiency of the daughter hadrons is obtained directly from
dividing the raw observed spectra of $\pi^{\pm}$ and p($\bar{p}$ )
by their respective invariant spectra. Powerful consistency checks
on trigger bias and $K^{\pm}$ $dE/dx$ uncertainty are possible by
comparing the $K^{\pm}$ spectra from jet away-side triggers and
published minimum-bias-trigger results~\cite{starppStrange} with the
invariant spectra of $K_{S}^{0}\rightarrow\pi^{+}+\pi^{-}$ from our
hadronic trigger.

\subsection{Trigger enhancement and efficiency for showering hadrons}
\begin{figure}
\center{\includegraphics[width=0.9\textwidth]{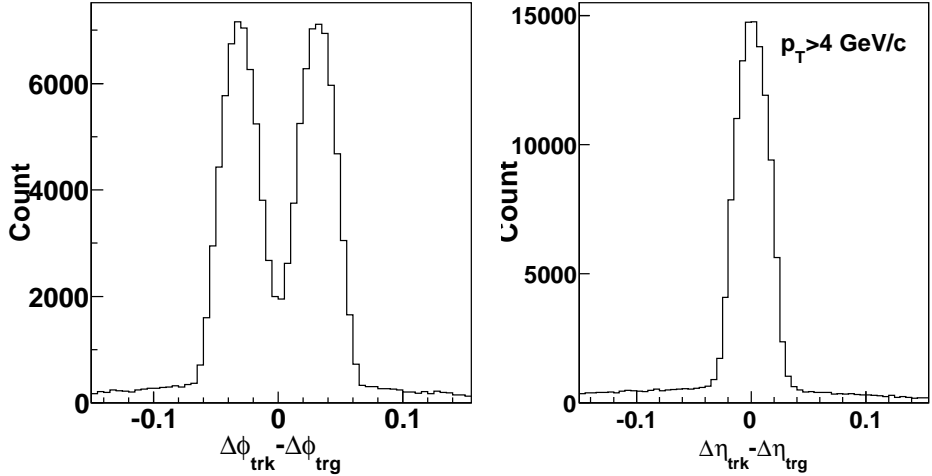}}
\caption{The $\Delta\phi$ and $\Delta\eta$ between tracks and BEMC
triggered towers. We note that the splitting into two peaks in $\Delta\phi$
(i.e. in the bending plane)
is due to the fact that the TPC track helices are projected to the BEMC surface while
in reality the hadronic showers are on average deep within the BEMC.}\label{deltaphieta}
\end{figure}

In this section, we provide the detailed procedure of obtaining the
$\pi^{\pm}$ and p($\bar{p}$) trigger efficiencies when either is
associated with the high-tower that passes the online trigger
threshold, where these are daughters from resonance
($\rho^{0}\rightarrow\pi^{+}+\pi^{-}$) or V0
($K^0_{S}\rightarrow\pi^{+}+\pi^{-}$,
$\Lambda(\bar{\Lambda})\rightarrow p(\bar{p})+\pi^{-(+)}$) decays.
In offline analysis, a track reconstructed in the TPC is projected
to the surface of the BEMC and associated with a shower
reconstructed from the BEMC tower energies. The distances between
the center of the triggered tower and the track projections are
shown in Figure~\ref{deltaphieta}. We require $|\Delta\phi| < 0.075$
$rad$ and $|\Delta\eta| < 0.075$ for matched tracks. Projecting the
backgrounds under the peaks, we find that these cuts include
$\sim$3\% of accidental coincidences, most of which will be further
reduced by additional cuts.

\begin{figure}
\begin{minipage}[t]{6.8cm}
\center{\includegraphics[width=0.9\textwidth]{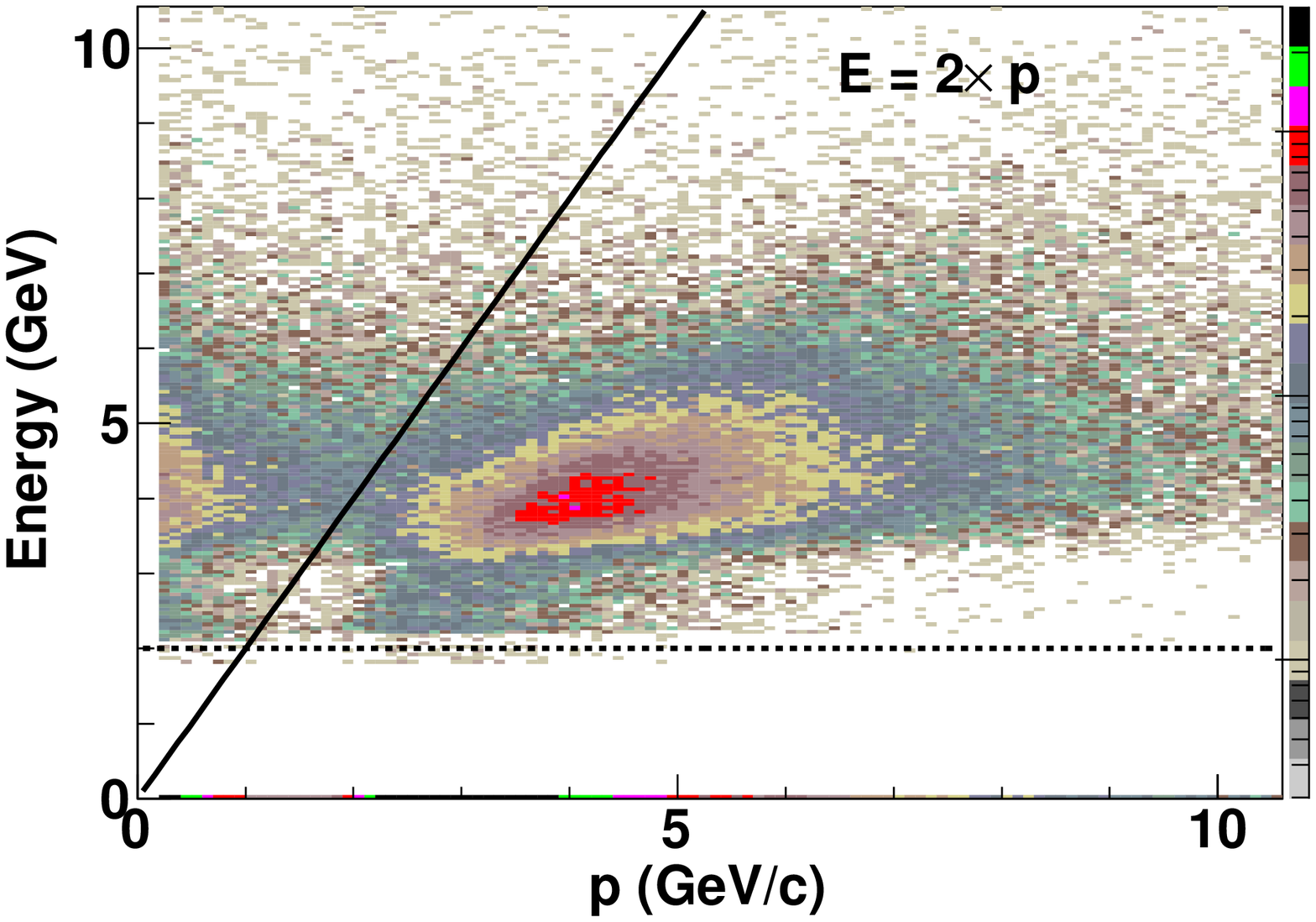}}
\caption{Energy of triggered towers versus momentum of matched
tracks. Only tracks with \pt~$>$3 GeV/c are used in the
analyses.}\label{Evsp}
\end{minipage}
\hspace{0.05in}
\begin{minipage}[t]{6.8cm}
\center{\includegraphics[width=0.9\textwidth]{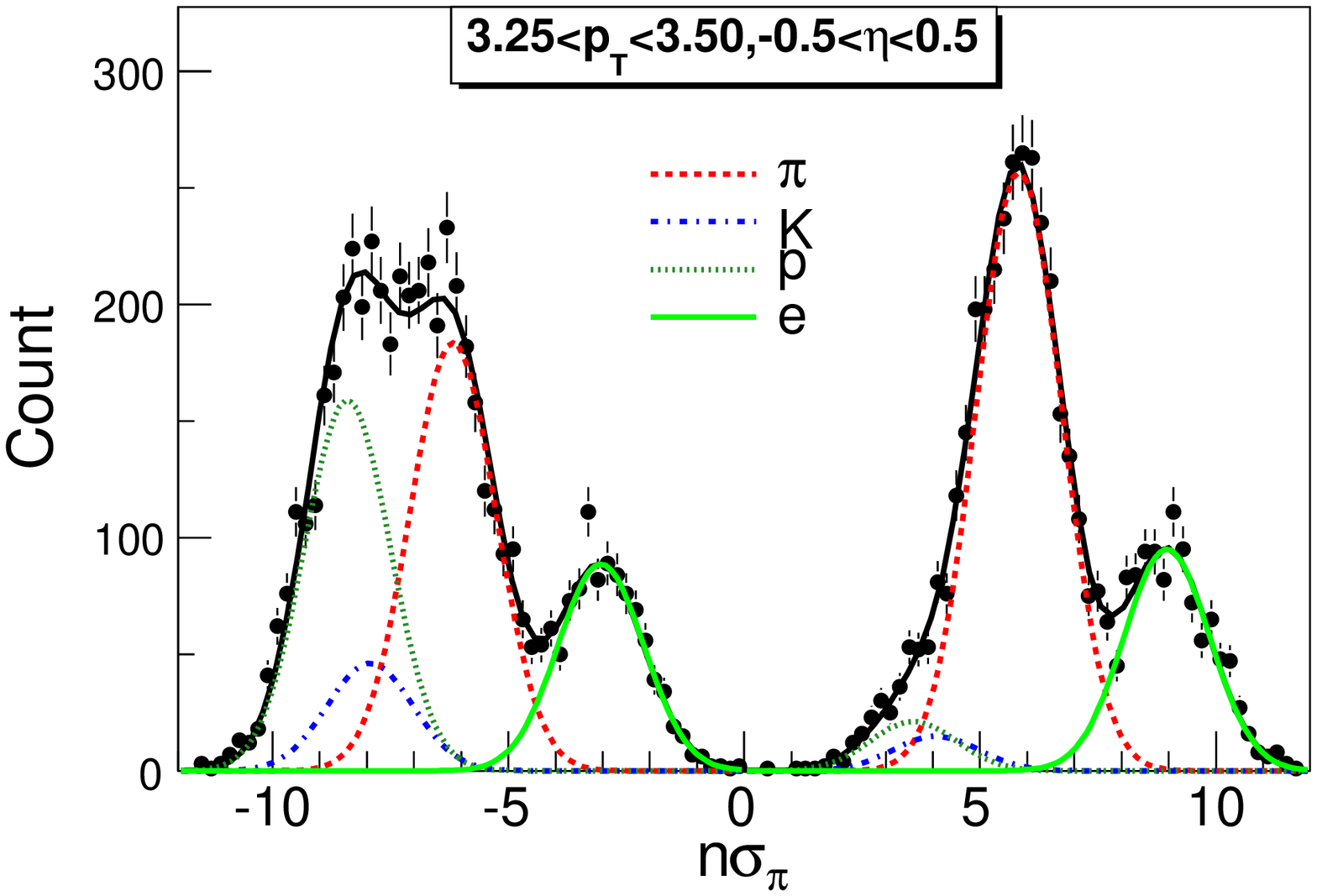}}
\caption{Normalized $dE/dx$ distributions at 3.25$<$\pt$<$3.50
$GeV/c$.}\label{nsigma}
\end{minipage}
\end{figure}

Although hadronic interactions in an electromagnetic calorimeter
develop showers for which much of the energy escapes the detector, a
significant fraction of hadrons leave a sizable captured energy
deposition. Figure~\ref{Evsp} shows the correlation between energy
deposited in the triggered tower and momentum of the matched track.
The particles are selected between two cuts (shown in the figure) on
$E \leq 2 \times p$, to remove accidental coincidences with
electromagnetic showers, and $E > 2$ $GeV$, to reject minimum ionizing
particles and other low energy background coincidences. This
provides a dataset with a wide momentum range for further particle
identification.

The normalized $dE/dx$,
n$\sigma_{\pi}$~\cite{dEdxCali,Ming,Bichsel,lndEdx} distributions at
$3.25 <$ \pt~$< 3.50$ $GeV/c$, offset by +6 for positive particles and
-6 for negative particles, are shown in Figure~\ref{nsigma}. The
peaks of triggered electrons and positrons are clearly separated
from charged hadrons. Statistics of charged pions and anti-protons
are significantly enhanced in comparison to the distributions for
minimum-bias-trigger data~\cite{dEdxCali}. The yield of triggered
anti-protons is much larger than that of protons because they
annihilate with the material in the BEMC and deposit an additional
$\sim$2 $GeV$ extra energy.

\begin{figure}
\begin{minipage}[t]{6.8cm}
\center{\includegraphics[width=0.9\textwidth]{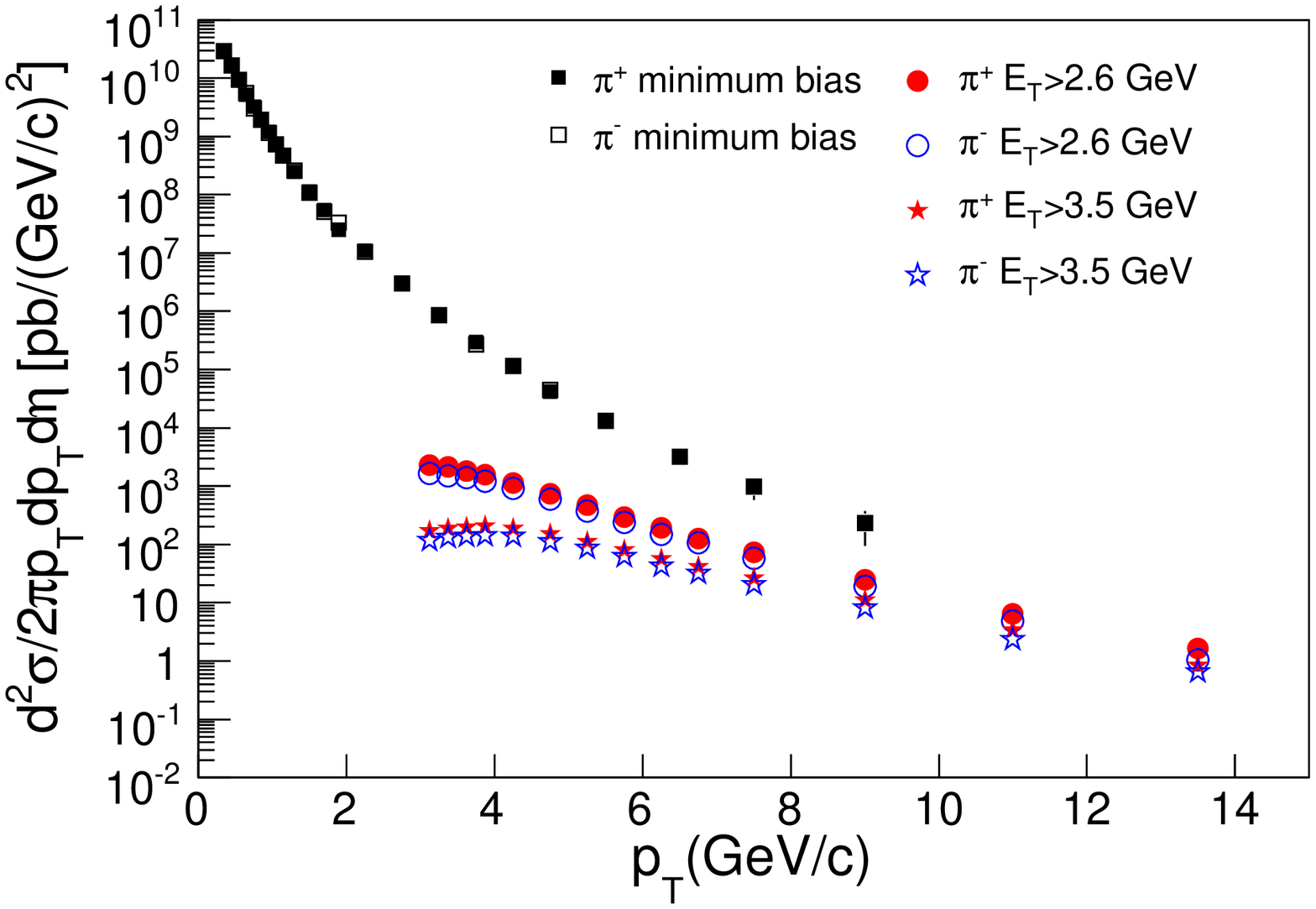}}%pionSpectraNew3.eps}}
\caption{pion \pt~spectra from minimum-bias events and BEMC-trigger events.}\label{pionSpectra}
\end{minipage}
\hspace{0.05in}
\begin{minipage}[t]{6.8cm}
\center{\includegraphics[width=0.9\textwidth]{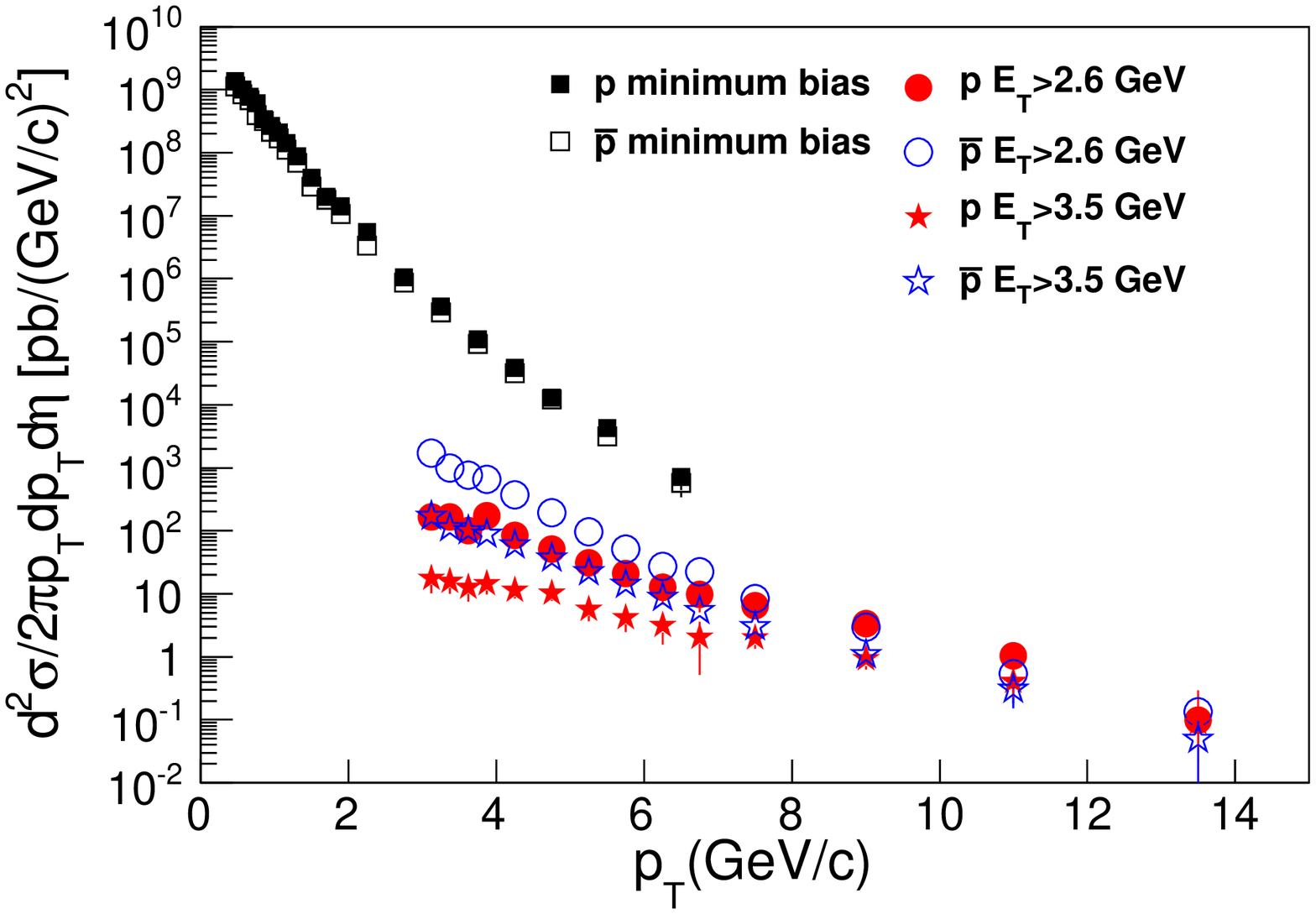}}%protSpectra_v3.eps}}
\caption{proton \pt~spectra from minimum-bias events and BEMC-trigger events.}\label{protonSpectra}
\end{minipage}
\end{figure}

\begin{figure}
\begin{minipage}[t]{6.8cm}
\center{\includegraphics[width=0.9\textwidth]{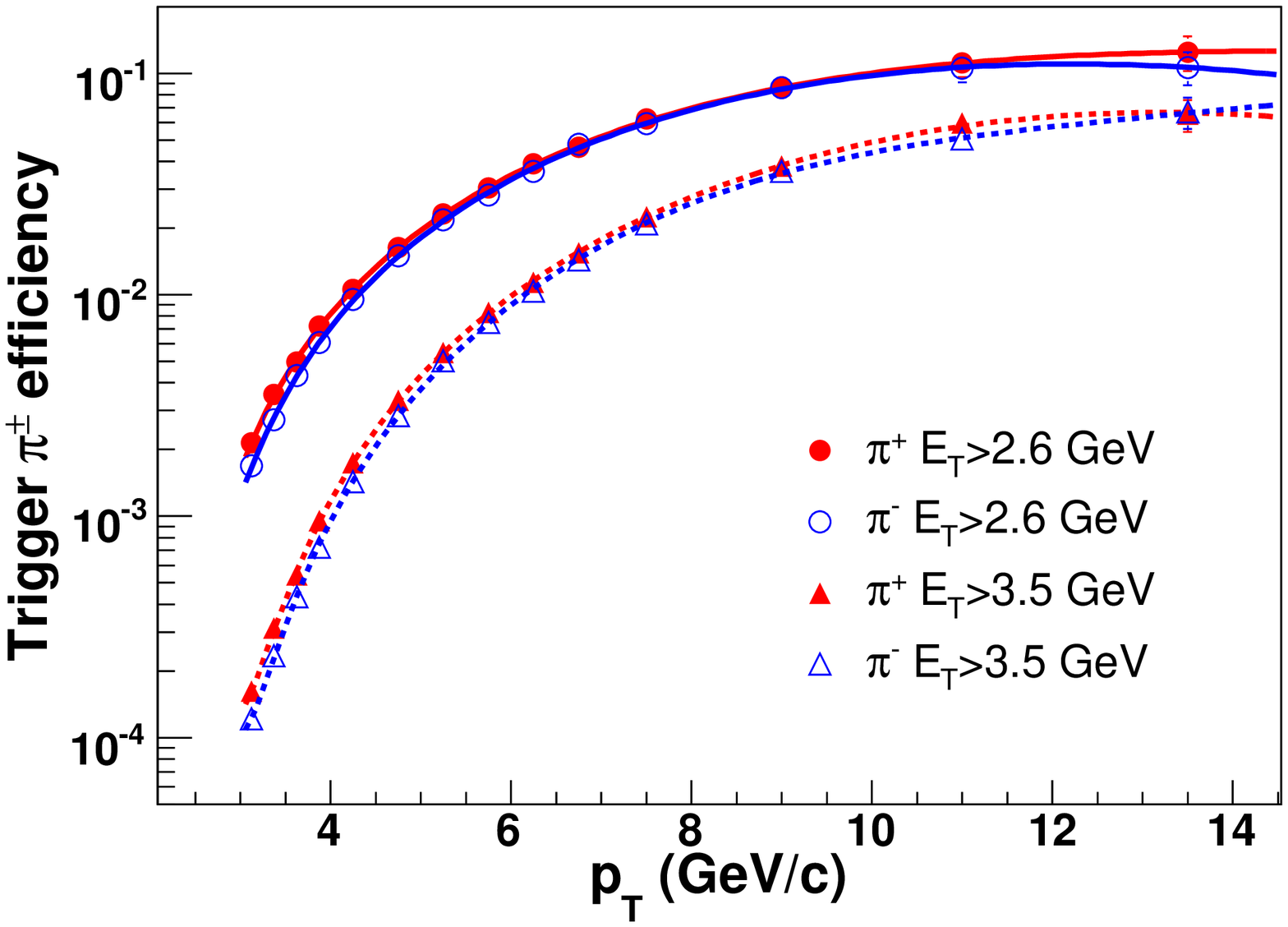}}%trigger6.eps}}
\caption{Trigger efficiency and tracking efficiency of pion from BEMC-trigger events.}\label{pionEfficiency}
\end{minipage}
\hspace{0.05in}
\begin{minipage}[t]{6.8cm}
\center{\includegraphics[width=0.9\textwidth]{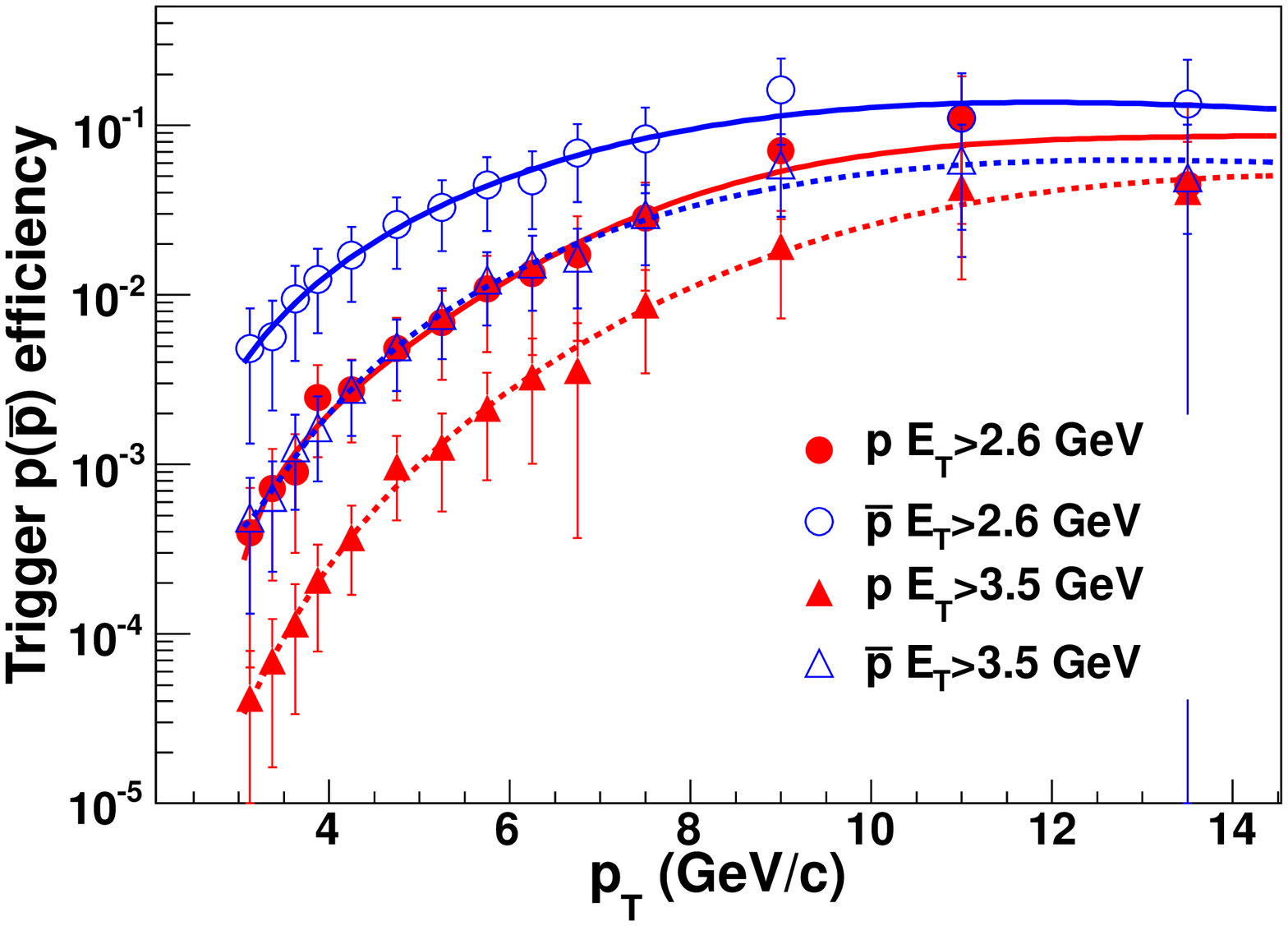}}%pbarpTrigger4paper.eps}}
\caption{Trigger efficiency and tracking efficiency of proton and
anti-proton from BEMC-trigger events.}\label{protonEfficiency}
\end{minipage}
\end{figure}

Efficiencies can be derived by dividing the raw \pt~spectra in the
BEMC-trigger events by the inclusive invariant spectra obtained
previously (shown in Figure~\ref{pionSpectra} and
Figure~\ref{protonSpectra} respectively), and these efficiencies for
pions and (anti-)protons are shown in Figure~\ref{pionEfficiency}
and Figure~\ref{protonEfficiency}. Although the efficiency is not as
high as a pure electromagnetic shower, the trigger enhancement is
quite high. Taking pion efficiency as an example, at \pt=5 $GeV/c$,
the triggered pion efficiency in HT1 is $\sim$2$\%$, and therefore
the 5.1 million HT1 events (from 0.65 $pb^{-1}$ sampled luminosity)
are equivalent to luminosity (0.65
$pb^{-1}$)$\times$$\sigma^{NSD}_{pp}$(30 $mb$)$\times$[trigger
efficiency](2$\%$)/[tracking efficiency](90$\%$) = $\sim$450 million
minimum-bias-trigger (non-single diffractive [NSD]) events. This
means a factor of $\sim$100 times more statistics for charged pions
at this \pt~ in this data sample than in the previously published
minimum-bias-trigger sample~\cite{starPIDpapers}. At higher \pt, the
trigger efficiency and the minimum-bias event equivalents are much
higher. The details can be found in Tables~\ref{efactor4pion}
and~\ref{efactor4prot}. Therefore, the BEMC-trigger data samples
significantly enhance the available statistics of the particles at
high \pt.

\begin{table}
\begin{center}
\caption{The number of equivalent minimum-bias events ($N_{eq}$) for
charged pion at given \pt~bin from 5.1 million HT1 and 3.4 million
HT2 events respectively.}\label{efactor4pion}

\begin{tabular}{|c|c|c|c|c|p{6.0cm}}
\hline \hline
\pt & $N_{eq}$($\pi^{+}$) HT1 & $N_{eq}$($\pi^{-}$) HT1 & $N_{eq}$($\pi^{+}$) HT2 & $N_{eq}$($\pi^{-}$) HT2\\
\hline
3.125  &4.66e+07  &3.61e+07  &1.53e+07  &1.16e+07 \\
\hline
3.375  &7.56e+07  &6.11e+07  &2.89e+07  &2.15e+07 \\
\hline
3.625  &1.11e+08  &9.28e+07  &5.29e+07  &4.10e+07 \\
\hline
3.875  &1.53e+08  &1.32e+08  &8.81e+07  &7.08e+07 \\
\hline
4.25  &2.29e+08  &2.03e+08  &1.64e+08  &1.37e+08 \\
\hline
4.75  &3.52e+08  &3.21e+08  &3.11e+08  &2.68e+08 \\
\hline
5.25  &4.97e+08  &4.64e+08  &5.15e+08  &4.54e+08 \\
\hline
5.75  &6.62e+08  &6.30e+08  &7.77e+08  &6.97e+08 \\
\hline
6.25  &8.41e+08  &8.11e+08  &1.10e+09  &9.98e+08 \\
\hline
6.75  &1.03e+09  &1.00e+09  &1.47e+09  &1.35e+09 \\
\hline
7.5  &1.32e+09  &1.30e+09  &2.13e+09  &1.97e+09 \\
\hline
9  &1.86e+09  &1.83e+09  &3.59e+09  &3.38e+09 \\
\hline
11  &2.38e+09  &2.26e+09  &5.33e+09  &5.02e+09 \\
\hline
13.5  &2.73e+09  &2.32e+09  &6.38e+09  &5.92e+09 \\
\hline \hline
\end{tabular}
\end{center}
\end{table}

\begin{table}
\begin{center}
\caption{The number of equivalent minimum-bias events ($N_{eq}$) for
proton and anti-proton at given \pt~bin from 5.1 million HT1 and 3.4
million HT2 events respectively.}\label{efactor4prot}
\begin{tabular}{|c|c|c|c|c|p{6.0cm}}
\hline \hline
\pt & $N_{eq}$(p) HT1 & $N_{eq}$($\overline{p}$) HT1 & $N_{eq}$(p) HT2 & $N_{eq}$($\overline{p}$) HT2\\
\hline
3.125  &7.88e+06  &9.66e+07  &3.75e+06  &4.41e+07 \\
\hline
3.375  &1.59e+07  &1.41e+08  &6.79e+06  &6.70e+07 \\
\hline
3.625  &2.53e+07  &1.94e+08  &1.17e+07  &1.04e+08 \\
\hline
3.875  &3.66e+07  &2.55e+08  &1.90e+07  &1.56e+08 \\
\hline
4.25  &5.77e+07  &3.62e+08  &3.53e+07  &2.63e+08 \\
\hline
4.75  &9.61e+07  &5.32e+08  &7.01e+07  &4.63e+08 \\
\hline
5.25  &1.50e+08  &7.30e+08  &1.25e+08  &7.29e+08 \\
\hline
5.75  &2.24e+08  &9.53e+08  &2.06e+08  &1.06e+09 \\
\hline
6.25  &3.19e+08  &1.19e+09  &3.18e+08  &1.45e+09 \\
\hline
6.75  &4.38e+08  &1.44e+09  &4.69e+08  &1.89e+09 \\
\hline
7.5  &6.56e+08  &1.81e+09  &7.77e+08  &2.62e+09 \\
\hline
9  &1.16e+09  &2.46e+09  &1.68e+09  &4.09e+09 \\
\hline
11  &1.66e+09  &2.92e+09  &3.20e+09  &5.50e+09 \\
\hline
13.5  &1.86e+09  &2.85e+09  &4.53e+09  &5.86e+09 \\
\hline \hline
\end{tabular}
\end{center}
\end{table}

\subsection{Resonance and V0 reconstruction}
The BEMC-trigger data sample not only increases the stable hadron
yields to tape, but also provides those high-statistic stable
hadrons for the resonance and V0 reconstructions. To reconstruct
\Ks~or \La(\aLa) via their dominant weak decay channels,
$K_S^{0}\rightarrow\pi^{+}+\pi^{-}$, $\Lambda(\overline{\Lambda})
\rightarrow p(\overline{p})+\pi^{-(+)}$, we look for at least one of
the decay daughters to be the particle firing the BEMC trigger. This
procedure has also been used in the cross section measurement
reported in Ref.~\cite{starppStrange,Topology}.

\begin{figure}
%\center{\includegraphics[width=0.9\textwidth]{v0Plots45_50.eps}}%v0Plots5_10.eps}}
\center{\includegraphics[width=0.99\textwidth]{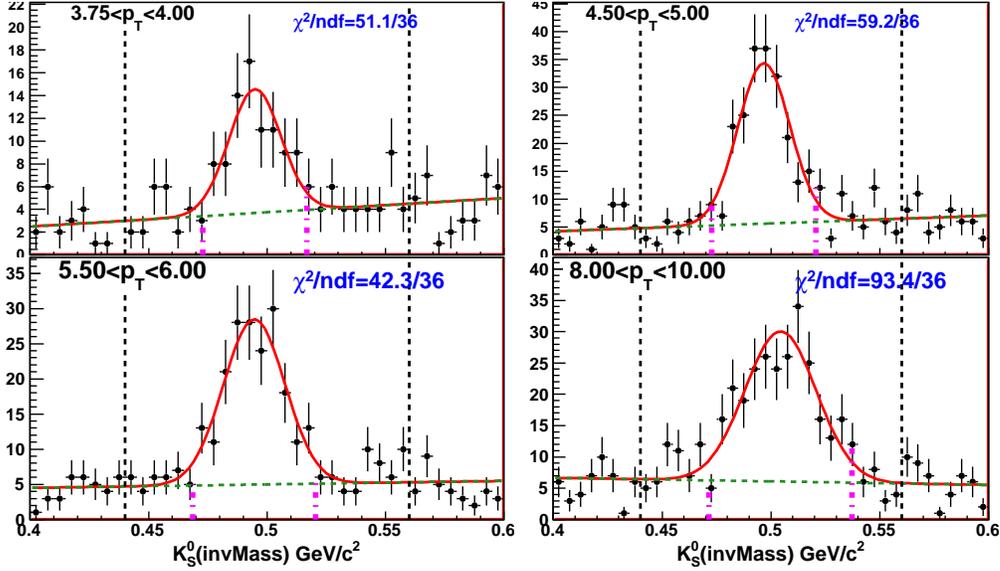}}
\caption{Invariant mass of \Ks~at several high
\pt~bins.}\label{V0Mass}
\end{figure}

\begin{figure}
%\center{\includegraphics[width=0.9\textwidth]{v0Plots45_50.eps}}%v0Plots5_10.eps}}
\center{\includegraphics[width=0.99\textwidth]{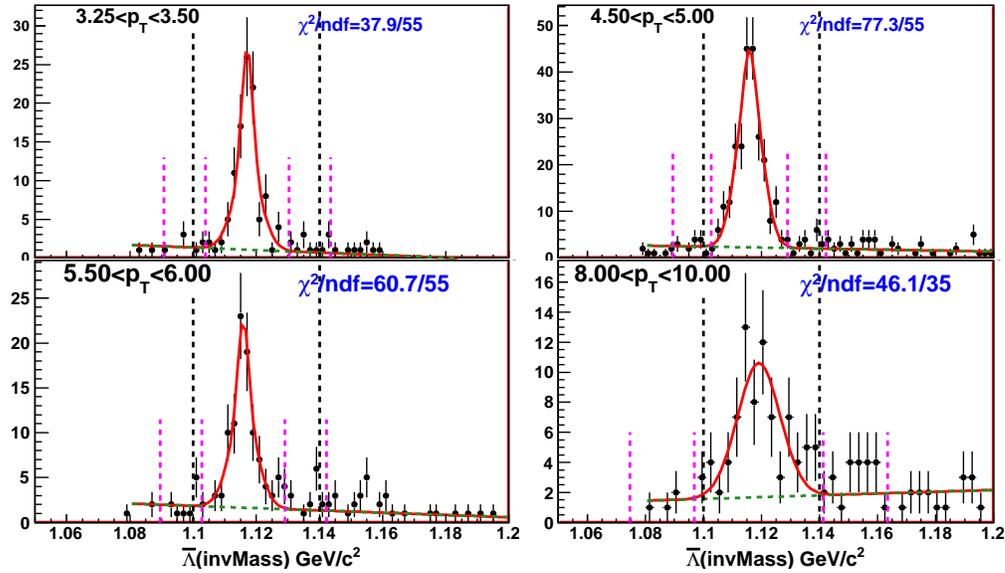}}
\caption{Invariant mass of \aLa~at several high
\pt~bins.}\label{V0MassLbar}
\end{figure}

The reconstructed event vertex is required to be along the beam axis
and within 100 $cm$ of the TPC center to ensure uniform tracking
efficiency.  A search is made in each event to find a (anti-)proton
and pion tracks of the opposite curvature. The tracks are then
paired to form a \Ks~or \La(\aLa) candidate and topological
selections are applied to reduce backgrounds. Figures~\ref{V0Mass}
and~\ref{V0MassLbar} show invariant mass distributions of the
triggered \Ks~and \La(\aLa) at high \pt~with only 3.4 million HT2
BEMC-trigger events, while 10 million minimum-bias events can only
reach 5 $GeV/c$ due to limited statistics~\cite{starppStrange}.

%{\bf can you replace these figures by:
%$http://www.star.bnl.gov/protected/lfspectra/xuyichun/webs/pphighptPIDpaper/supportdocument/trigger/para13_Ht1lw.gif$
%and new plots from Da Hongyu?}

\begin{figure}
\begin{minipage}[t]{6.8cm}
\center{\includegraphics[width=0.9\textwidth]{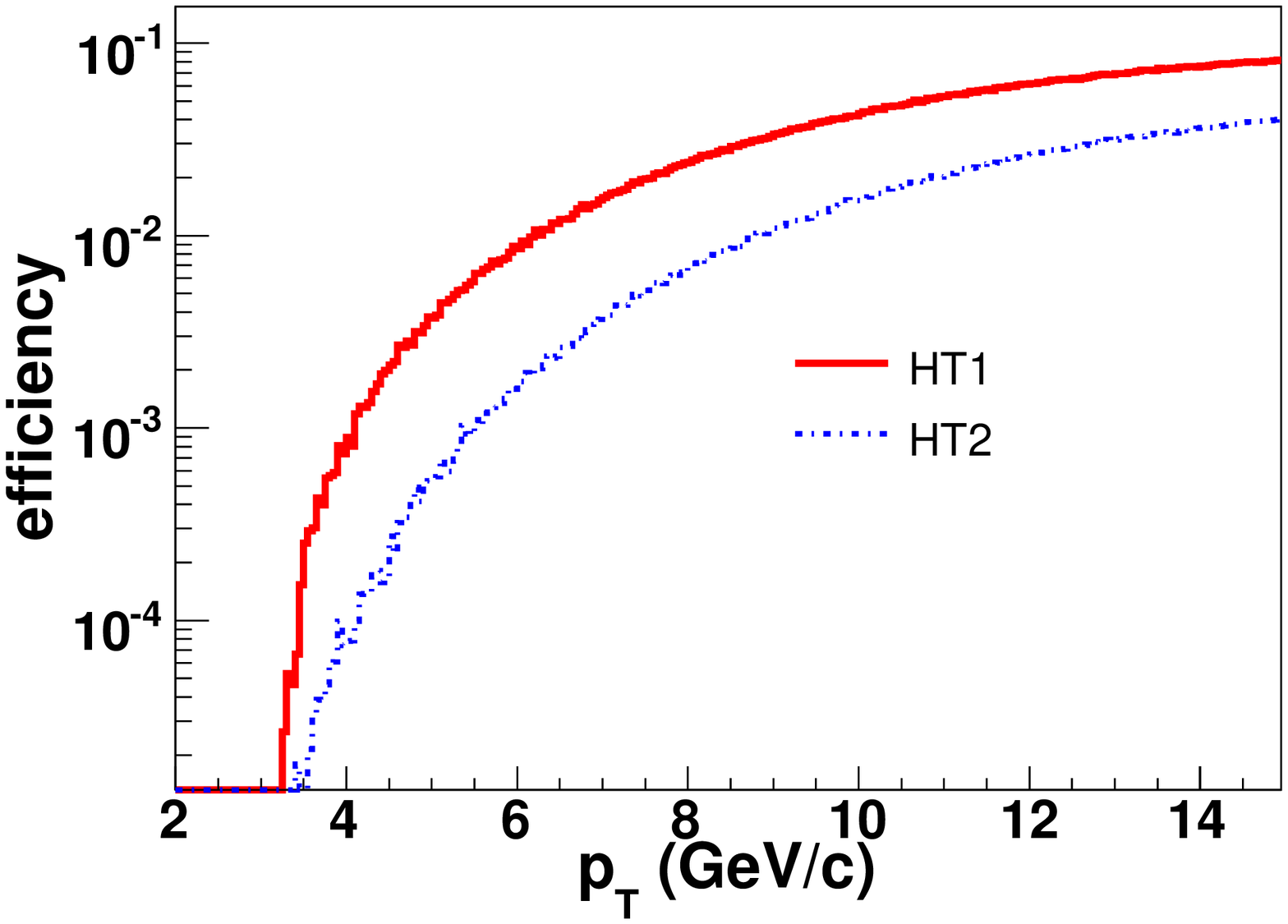}}%effKshort.eps}}
\caption{Kinematic efficiency of \Ks~vs. \pt.}\label{KsEff}
\end{minipage}
\hspace{0.05in}
\begin{minipage}[t]{6.8cm}
\center{\includegraphics[width=0.9\textwidth]{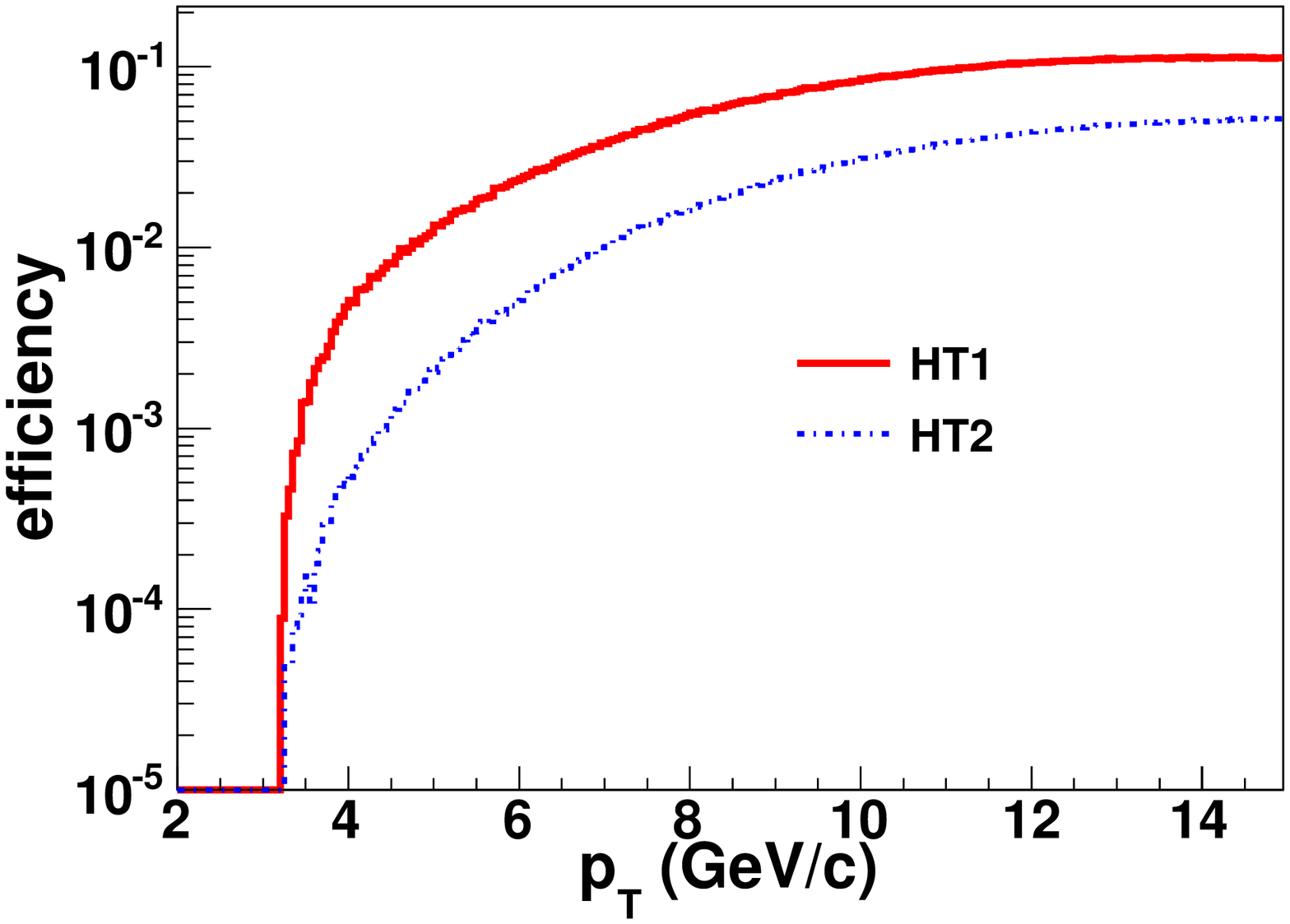}}%plot4ALambda.eps}}
\caption{Kinematic efficiency of \aLa~vs. \pt.}\label{LambdaEff}
\end{minipage}
\begin{minipage}[t]{6.8cm}
\center{\includegraphics[width=0.9\textwidth]{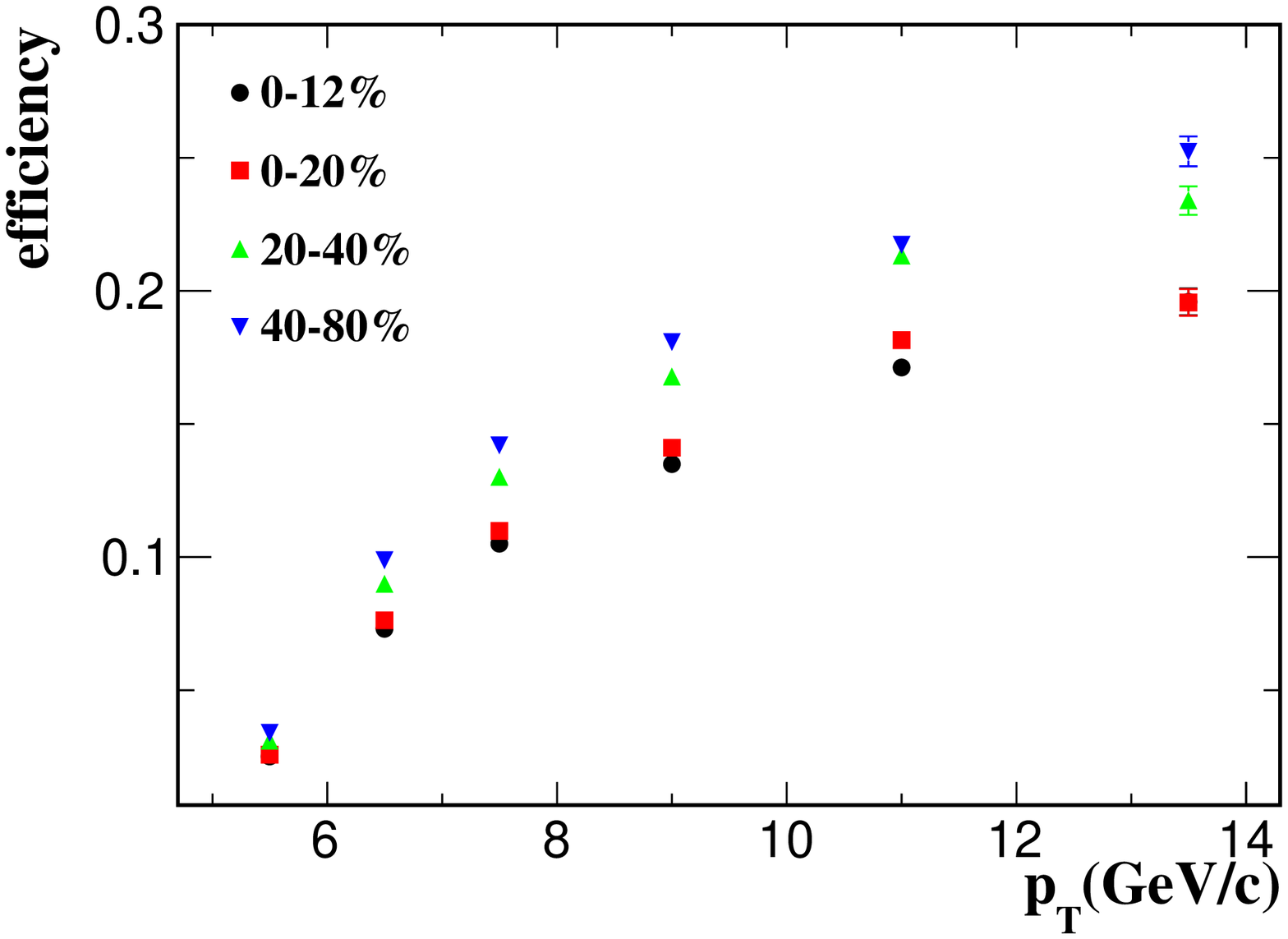}}
\caption{Kinematic efficiency of $\rho^{0}$ vs. \pt.}\label{rho0Eff}
\end{minipage}
\end{figure}

To obtain the invariant spectra, we need to apply the correction
factors due to the efficiencies of trigger, tracking and topological
cuts. The correction is divided into two factors: kinematic
efficiency and topological efficiency. We define the kinematic efficiency
to include the effects of the BEMC response and trigger, the TPC
tracking efficiency, and the acceptance due to kinematics. The
topological efficiency includes the effects due to the topological
requirements in V0 reconstruction and is found to be $p_T$
independent at 92\% for \Ks~(the topological
efficiencies of $\Lambda(\overline{\Lambda})$ are still under study
as of this writing)~\cite{starppStrange}. Figures~\ref{KsEff},~\ref{LambdaEff}
and~\ref{rho0Eff} are the kinematic efficiencies as a function of
\pt~for the reconstruction of parent particles \Ks~, $\bar{\Lambda}$
and $\rho^{0}$, respectively.

\begin{figure}
\center{\includegraphics[width=0.9\textwidth]{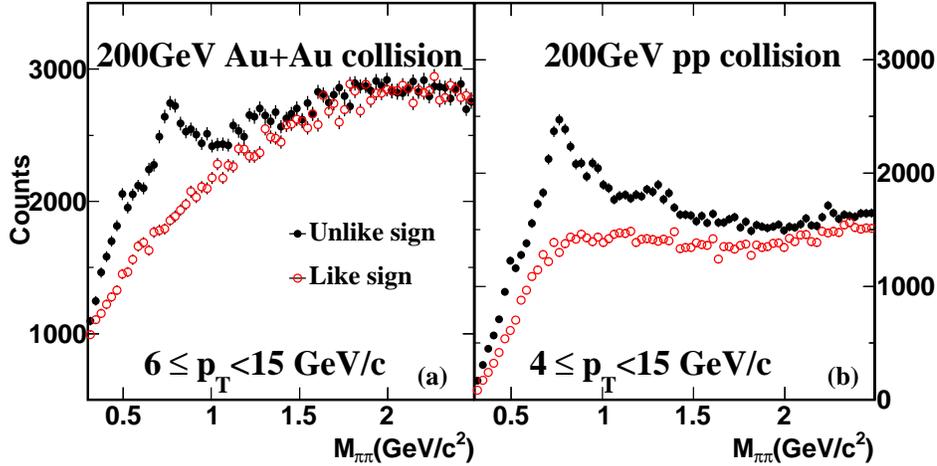}}
\caption{Invariant mass distributions of the unlike-sign pairs $\pi^{+}+\pi^{-}$
and like-sign pairs $\pi^{\pm}+\pi^{\pm}$ around
the $\rho^{0}$ mass. Panels (a) and (b) show the distributions from
Au+Au and $p+p$ collisions respectively.}\label{pi2raw}
\end{figure}

\begin{figure}
\center{\includegraphics[width=0.9\textwidth]{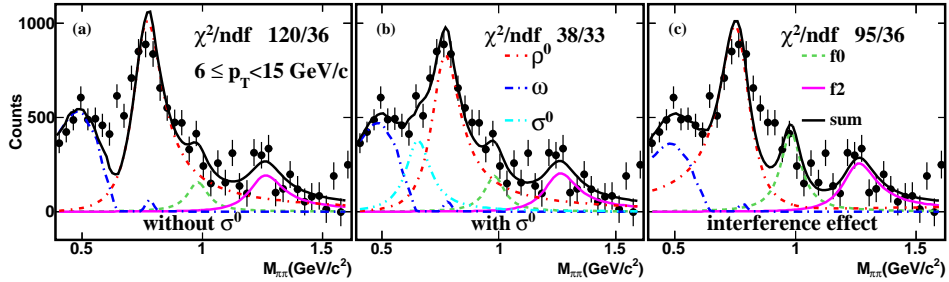}}
\caption{Invariant mass distribution of $\pi^{+}+\pi^{-}$ around
the $\rho^{0}$ mass. The three panels show fits for (a) a four-species
cocktail, (b) cocktail with additional $\sigma_0$, and (c)
cocktail with an additional interference term.}\label{rho0mass}
\end{figure}

Resonances ($\rho^{0}\rightarrow\pi^{+}+\pi^{-}$, $K^{*}\rightarrow
K^{\mp}+\pi^{\pm}$ etc.) can be reconstructed in a similar fashion.
Since $\rho^{0}$ decays strongly with a lifetime of about 1 $fm/c$,
reconstruction from its pion daughter pairs is made without the
displaced decay topological constraints used in V0 reconstruction.
The invariant masses of unlike-sign ($\pi^{+}+\pi^{-}$) and
like-sign ($\pi^{\pm}+\pi^{\pm}$) pion pairs are calculated and
shown in Figure~\ref{pi2raw}. Panel (a) from this figure is for the
invariant distributions in Au+Au collisions while panel (b) is for
\pp~collisions. The like-sign pairs represent the random
combinatoric background and the excess above this background
distribution in the unlike-sign is attributable to particle decays
($\rho^{0}$, $\omega$, $f_{0}$, and $f_2$). Figure~\ref{rho0mass}
shows the $\pi^{+}+\pi^{-}$ invariant mass distribution after
like-sign background subtraction.

For the line shape of $\rho^{0}\rightarrow\pi^{+}+\pi^{-}$, the
procedure and formula from a low-$p_T$ study are used with the
$\rho^{0}$ mass at 775 $MeV/c^2$ and Breit-Wigner width of 155
$MeV/c^2$~\cite{starrho}. As with that study, a four-species
cocktail describes the data quite well except that it
under-describes data at invariant mass around 600 $MeV/c^{2}$. This
is clearly visible in Figure~\ref{rho0mass}(a). To investigate the
possible missing components of the cocktail and how they impact the
extracted $\rho^0$ yields, we perform two additional studies by
adding a $\sigma^0$ to the cocktail and by adding an interference
term between $\rho^0\rightarrow\pi^{+}+\pi^{-}$ and direct
$\pi^{+}+\pi^{-}\rightarrow\pi^{+}+\pi^{-}$ scattering. Inclusion of
the possible $\sigma^{0}$ particle~\cite{sigma0:11} (mass at
$\sim$600 $MeV/c^2$ and Breit-Wigner width scanning from 100 to 500
$MeV/c^2$) results in 20\% lower $\rho^{0}$ yields and improves the
$\chi^{2}$ per degree of freedom ($\chi^{2}/NDF$) from 120/36 to
38/33, a factor of nearly 3 improvement. This fit is shown in
Figure~\ref{rho0mass}(b) and is used to obtain the default
$\rho^{0}$ yields, where the $\sigma^{0}/\rho^{0}$ ratio is about
25\% independent of $p_T$. An additional systematic check is
performed using the modified Soeding parametrization for a possible
interference effect on the $\rho^{0}$ line shape~\cite{UPCrhopaper}.
This parametrization and the relative amplitudes of its two
interference terms are determined from
clear signals and well-defined processes in ultra-peripheral Au+Au
collisions. Both the resulting $\chi^{2}/NDF$ and the $\rho^{0}$
yield fall between the other two fits. Figure~\ref{rho0mass}(c)
shows that including the interference over-corrects the high-mass tail of the
$\rho^0$ spectral distribution and therefore under-predicts the
data.

\section{Summary}
An electromagnetic calorimeter (the STAR BEMC) is leveraged to
enhance the event sample containing charged hadrons at high
transverse momentum, utilizing the STAR TPC for momentum
reconstruction and species identification through $dE/dx$. The
away-side hadrons opposite triggered jet patches (high collective
energy in a group of BEMC towers) are used to determine inclusive spectra of
identified stable charged hadrons at high \pt. Events triggered by high energy in a
single BEMC tower are used to find hadronic shower candidates by
matching these towers to high \pt~TPC tracks. These tracks are then
paired with other charged hadrons to reconstruct \Ks~and \La(\aLa)
through their dominant decay channels:
\Ks$\rightarrow\pi^{+}+\pi^{-}$ and $\Lambda(\overline{\Lambda})
\rightarrow p(\overline{p})+\pi^{-(+)}$. With this method, spectra
of the identified charged hadrons, \Ks~and \La(\aLa) have been extended
to higher \pt~($\sim$12 $GeV/c$) under the existing detector and
RHIC luminosity capabilities~\cite{starppStrange}. This method is
also used to extend the \pt~reach for efficiently reconstructing
strongly decaying particles, such as $\rho^0$ and $K^{\star}$.

\section*{Acknowledgments}
We thank the STAR Collaboration, the RHIC Operations Group and RCF
at BNL, and the NERSC Center at LBNL for their support. This work
was supported in part by the Offices of NP and HEP within the U.S.
DOE Office of Science; Authors Yichun Xu and Zebo Tang are supported
by National Natural Science Foundation of China under Grant No.
11005103 and No. 11005104, and Hongyu Da and Xiangli Cui are
supported by the Knowledge Innovation Program of the Chinese Academy
of Sciences, Grant No. kjcx2-yw-a14, NSFC(10620120286, 10620120285).

%part by NSFC 10475071, National Natural Science
%Foundation of China  under Grant No. 10610286 (10610285) and
%Knowledge Innovation Project of Chinese Academy of Sciences under
%Grant No. KJCX2-YW-A14.

%\newpage

% \appendix

% \section{}
% \label{}

\end{document}